\documentclass[amssymb,twocolumn,showpacs,prl,aps]{revtex4}
\usepackage{graphicx}
\begin{document}

\title{Hard sphere-like dynamics in a non hard sphere liquid
}

\author{T. Scopigno$^{1,2}$, R. Di Leonardo$^{1,2}$, L. Comez$^{1,3}$, A.Q.R. Baron$^4$, D. Fioretto$^{1,5}$, G. Ruocco$^{1,6}$}

\affiliation{
$^{1}$INFM CRS-SOFT, c/o Universit\'a di Roma "LaSapienza",~I-00185, Roma, Italy\\
$^{2}$INFM UdR-RS, c/o Dipartimento di Fisica, Universit\'a di Roma "La Sapienza", I-00185, Roma, Italy\\
$^{3}$INFM UdR-PG, c/o Dipartimento di Fisica, Universit\'a di Perugia, via Pascoli, I-06123 Perugia, Italy\\
$^{4}$SPring-8/JASRI, 1-1-1 Kouto, Mikazuki-cho, Sayo-gun,
Hyogo-ken 679-5198 Japan.\\
$^{5}$Dipartimento di Fisica, Universit\'a di Perugia, via Pascoli, I-06123 Perugia, Italy\\
$^{6}$Dipartimento di Fisica, Universit\'a di Roma ``La
Sapienza'', I-00185, Roma, Italy.}
\date{\today}

\begin{abstract}

The collective dynamics of liquid Gallium close to the melting
point has been studied using Inelastic X-ray Scattering to
probe lengthscales smaller than the size of the first coordination shell.
Although the structural properties of this partially covalent
liquid strongly deviate from a simple hard-sphere model, the
dynamics, as reflected in the quasi-elastic scattering, are
beautifully described within the framework of the extended heat
mode approximation of Enskog's kinetic theory, analytically
derived for a hard spheres system. The present work
demonstrates the applicability of Enskog's theory to non hard-
sphere and non simple liquids.

\end{abstract}


\maketitle

Since the first appearance of Enskog's theory \cite{ens_hs},
special attention has been devoted to the theoretical and
numerical study of the dynamics of hard spheres fluid as a useful
tool to mimic the behavior of simple liquids
\cite{lebo_hs,furt_hs}. During the seventies, the development of
neutron scattering (INS) facilities provided a database of
dynamical properties of simple liquids, while the enormous
advances of computational capabilities allowed recourse to the
hard sphere model to evaluate transport coefficients
\cite{all_hs1} and neutron scattering response \cite{all_hs}. At
the same time, on the theoretical side, the dynamical properties
of an ensemble of hard spheres have been investigated by
revisiting Enskog's original kinetic theory. One of the most
significant outcomes of these efforts is the so called extended
hydrodynamic theory \cite{desh_hyd0,desh_hs,coh_hs1,kag_hs}, which
has been particularly successful in describing INS data
\cite{coh_hs}. While Enskog's kinetic theory, strictly speaking,
applies to hard spheres fluids only, it has the advantage of
readily predicting the $Q$-dependence of some transport
coefficients, as opposed to more rigorous and involved theories
using memory functions or generalized hydrodynamics. Thus this
theory remains a very useful tool.

The extent of validity of Enskog's kinetic approach has been
satisfactorily tested against inelastic neutron scattering data
collected in very simple liquids including Kr, Ar, Ne, Rb
\cite{coh_hs}.  These are all similar in that their structure is
well described by a hard sphere model, with an effective density
and radius which can be determined by matching the first peak of
the static structure factor. However, no attempt has been made, to
our knowledge, to verify the applicability of the theory to
liquids markedly deviating from hard-sphere like structures,
probably because very few accurate, constant $Q$, experimental
determination of coherent spectra exist in these systems
\cite{dal_bi}.

In this letter, we present a study of the collective dynamic
structure factor ($S(Q,\omega$)) at wavevectors beyond the first
maximum of the static structure factor ($S(Q)$), using inelastic
x-ray scattering. The system we address is liquid gallium, as the
purpose of this work is to ascertain validity of the connection
between structural and dynamical properties as predicted by
Enskog's hard sphere theory for a liquid with strong non-hard-
sphere structural properties. Among simpler liquids, Ga has
peculiar structural and electronic properties. In addition to the
low melting temperature ($T_m=303$ K), it shows an extended
polymorphism in the solid phase with complex crystal structures
characterized by the competition between metallic and covalent
bindings. Despite the nearly free electron electronic DOS,
anomalies associated with some covalency residue have been
reported \cite{gong_ga}. Among them, the most important for the
present purpose is that the first peak of the $S(Q)$ presents a
hump characteristic of non close-packed liquid structures
\cite{asc_ga,bf}, which marks a significant departure from the
smoother behavior of the hard sphere structure factor. Although
the dynamics of liquid Gallium has been previously studied by INS
\cite{ber_ga1,ber_ga2,bov_ga} and IXS \cite{scop_prlga}, the
restricted available $Q$ range in the first case, and the presence
of incoherent scattering in the second case, prevented a study of
the collective dynamics in a $Q$ region beyond the main peak of
the structure factor like the present one.

We demonstrate that the collective dynamics of molten gallium is
well described by Enskog theory up to wavevectors as large as
three times that of the first maximum of the static structure
factor.  While the reduced density and mean free path of liquid
Gallium at melting point fall into the region of applicability of
Enskog's theory, the structural peculiarities of this system would
discourage any recourse to predictions stemming from an hard
sphere paradigm. However, the hard-sphere model turns out to
describe the dynamics even in the region where the structure
factor is notably different than a hard-sphere model.

The experiment was performed at the high resolution inelastic
scattering beam-line (BL35XU) \cite{bar_spring8} of SPring-8 in
Hyogo prefecture, Japan. High resolution was obtained using the (9
9 9) reflection of perfect silicon crystals while a backscattering
geometry ($\frac{\pi}{2} - \theta_B \approx 0.3$ mrad) was used
(for both monochromator and analyzers) in order to obtain large
angular acceptance. The flux onto the sample was $\approx
10^{10}$ photons/sec (100 mA electron beam current) in a 1.8 meV
bandwidth at 17.793 keV. The use of 4 analyzers crystals, placed
with 0.70 degree spacing on the $10$ m two-theta arm (horizontal
scattering plane), and 4 independent detectors, allowed collection
of 4 momentum transfers simultaneously. Slits in front of the
analyzer crystals limited their acceptance to 0.24 nm$^{-1}$ in
the scattering plane.  The over-all resolution of the spectrometer
was about 2.8 meV, depending on the analyzer crystal. Typical data
collection times were 200 s/bin, where the bin size was fixed at
$0.25$ meV.

The Ga sample, about 80 microns in thickness, was placed in sample
cell with thin (2 $\times$ 250 micron) single crystal sapphire
windows. This was held, in vacuum, at a constant temperature of
315 K.

\begin{figure} [h]
\centering
\includegraphics[width=.45\textwidth]{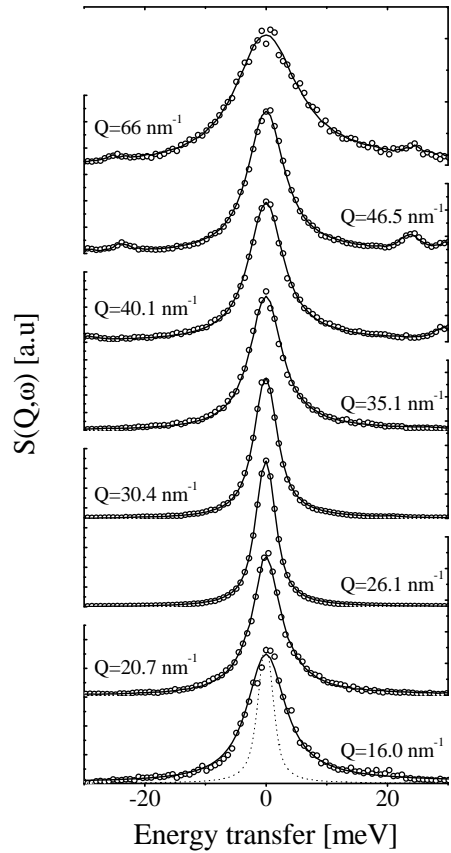}
\vspace{-.01cm}\caption{IXS spectra of liquid Ga ($T=315$ K) at
the indicated fixed $Q$ values (open dots). Also reported is the
instrumental resolution (which is the same -dotted line- for all
the reported spectra) and the best-fit lineshapes (continuous
line, see text). The structures at larger energy transfers are
phonons from the sapphire windows of the sample container.}
\label{panel}
\end{figure}

Measured IXS spectra are reported in Fig. \ref{panel} for selected
constant values of momentum transfer $Q$. The strong central peak
is readily visible along with some background due to phonons in
the sapphire windows. Solid lines in the figure are best-fits with
two components: \textit{i)} a single Lorentzian line, modified in
the usual way to account for detailed balance \cite{scop_jpc}, and
convolved with the instrument resolution and \textit{ii)} the
empty cell background measured at each Q, normalized by the sample
transmission. As it will be shown in the following, the Lorentzian
spectrum is predicted by Enskog's kinetic theory, which also
furnishes explicit expressions for the $Q$-dependence of the
parameters describing its linewidth and amplitude (see, for
instance, Eq. (\ref{zh_hs})).

\begin{table*}[h]
\centering
\renewcommand{\arraystretch}{0.5}
\begin{tabular}{||c|c||c|c|c|c|c|c||} \hline\hline $\sigma$&$D_E$&$\chi ^2$&$\sigma$ [nm$^{-1}$]&$\hbar D_E$ [meV/nm$^{-2}$]&Packing fraction&Reduced density&$l_E$ [nm]\\
&&&&&$\varphi=\frac{\pi}{6} \rho \sigma^3$&$V_0/V=\rho \sigma^3/\sqrt 2$& \\
\hline
 FREE&FREE&46&0.279&1.14$\times10^{-3}$&0.601&0.811&9.32$\times 10^{-3}$\\ \hline
 FREE&Eq.(\ref{DE})&108&0.265&1.13$\times 10^{-3}$&0.513&0.693&9.24$\times 10^{-3}$\\ \hline
 MP&FREE&353&0.246&1.09$\times10^{-3}$&0.411&0.555&8.90$\times 10^{-3}$\\ \hline
 MP&Eq.(\ref{DE})&4988&0.246&2.18$\times10^{-3}$&0.411&0.555&1.78$\times 10^{-2}$\\ \hline
 WA&FREE&681&0.226&9.54$\times10^{-4}$&0.318&0.430&7.78$\times 10^{-3}$\\ \hline
 WA&Eq.(\ref{DE})&38533&0.226&3.78$\times10^{-3}$&0.318&0.430&3.08$\times 10^{-2}$\\ \hline
 \hline
\end{tabular}
\caption{Structural and dynamical properties of liquid gallium at
the melting point, as derived by the present experiment. Each row
is relative to a different fitting strategy, since the effective
hard sphere diameter has been determined: \textit{i)} as an
adjustable parameter, \textit{ii)} by the position of the main
peak of the structure factor (MP), \textit{iii)} as the value
corresponding to the hard sphere structure factor which better
approximates the measured one (WA). Correspondingly, the Enskog's
diffusion coefficient, $D_E$, can be obtained either as an
additional adjustable parameters or fixed through Eq.(\ref{DE}).
All the other quantities are derived by $\sigma$ and $D_E$
according to the definitions given in the text.} \label{table}
\end{table*}

The main idea behind Enskog's theory is to evaluate the
correlation functions of the microscopic quantities, such has the
density-density correlation function of interest here, replacing
the Liouville operator and the set of relevant dynamical variables
defined at the $N-$particle ensemble level, with the one particle
Enskog operator $L$ and appropriate single particle variables.
Within this framework, the dynamic structure factor reads
(\cite{desh_hyd0,desh_hs,coh_hs1,kag_hs}):

\begin{eqnarray}
S(Q,\omega)&=&S(Q)\frac{1}{\pi}\mathrm{Re}\left \{ \left \langle
\frac{1}{i\omega-L(\mathbf{Q},\mathbf{v}_1)}\right \rangle_1
\right \} \nonumber
\\ &=&S(Q)\frac{1}{\pi}\mathrm{Re} \left
\{ \sum_j \frac{B_j(Q)}{i\omega+z_j(Q)} \right \} \label{ehm}
\end{eqnarray}

with

\begin{equation}
B_j(Q)=\langle \Psi^{(L)}_j(\mathbf{Q},\mathbf{v}_1)\rangle_1
\langle \Psi^{(R)*}_j(\mathbf{Q},\mathbf{v}_1) \rangle_1
\label{hscoeff}
\end{equation}

The subscript $\langle ... \rangle_1$ indicates a single particle
averages over the Maxwell Boltzmann velocity distribution
function, while $z_j(Q)$, $\Psi^{(L)}_j(\mathbf{Q},\mathbf{v}_1)$
and $\Psi^{(R)}_j(\mathbf{Q},\mathbf{v}_1)$ are the eigenvalues
and the left and right eigenvectors of $L$, respectively.

There are several approaches to determine the spectrum of $L$, and
different approximations can be applied depending on the density
and kinematic regions of interest. Following Ref. \cite{kag_hs},
these regions are identified by the values of the reduced density

\begin{equation}
V_0/V=\rho \sigma^3/\sqrt 2 \label{reddens}
\end{equation}

in which $V_0$ is the closest packing volume for spheres of radius
$\sigma$ and number density $\rho$, and by Enskog's mean free path
$l_E=l_0/\chi$, with $l_0=1/\sqrt 2 \rho \pi \sigma^2$ the
Boltzmann mean free path and $\chi=g(\sigma)$ the pair
distribution function evaluated at the contact point between two
spheres.

Taking, as usual, density, momentum and energy as relevant
variables, the lower three eigenvalues of $L$ always go to zero
with $Q\rightarrow 0$. By introducing Enskog's thermal diffusion
coefficient $D_E$, the adiabatic sound velocity $c_o$ and the
sound damping coefficient $\Gamma_E$, it can be shown that these
low lying eigenvalues are \cite{desh_hs,coh_hs1}:

\begin{eqnarray}
&&z_1(Q)=z_h(Q)=D_{E}Q^2 \nonumber \\
&&z_{2,3}(Q)=z_\pm (Q)=\pm i c_0Q+\Gamma_E Q^2 \label{coeff_hyd}
\end{eqnarray}

\noindent

This limit is practically attained at low densities ($V_0/V <0.1$
and therefore $l_E\approx l_0$) and sufficiently small $Q$'s
($Q\sigma<<1$). This condition normally occurs in the case of
light scattering experiments from dilute gases ($Ql_0\approx 1$),
and one speaks in terms of three extended hydrodynamic modes.

In dense fluids ($V_0/V$ approximately larger than 0.35),
Kamgar-Parsi et al. have shown that a description in terms of
three \textit{effective} hydrodynamic modes still applies
\cite{kag_hs}, and the low $Q$ limit of these modes is again
coincident with the hydrodynamic result. At variance with the
previous case, however, the hydrodynamic scheme breaks down as
soon as $Ql_E \approx 0.05$. Above this value, Enskog's operator
is dominated by binary collisions and only the extended heat mode
is well separated by all the other modes \cite{kag_hs}. The
following approximate expression for the extended heat mode can be
given:

\begin{equation}
z_h(Q)=\frac{D_EQ^2}{S(Q)}d(Q) \label{zh_hs}
\end{equation}

\noindent in which $D_E$ is Enskog diffusion coefficient and

\begin{eqnarray}
d(Q)\approx \frac{1}{1-j_0(Q\sigma)+2j_2(Q\sigma)} \label{dq}
\end{eqnarray}

can be expressed in terms of the first two even order Bessel
spherical functions. Enskog's diffusion coefficient is related to
the Boltzmann diffusion coefficient

\begin{eqnarray}
D_0=\frac{3}{8\rho \sigma^2}\sqrt{\frac{k_B T}{\pi m}} \approx
\frac{0.216}{\rho \sigma^2}\sqrt{\frac{k_B T}{m}} \nonumber
\end{eqnarray}

\noindent by the collision enhancing term $g(\sigma)$ as
$D_E=D_0/g(\sigma)$. Plugging the well known analytic expression
of $g(\sigma)$ for hard spheres \cite{FABER} in the previous
equation, one easily gets a final expression for $D_E$ in terms of
the packing fraction $\varphi=\pi \rho \sigma^3 / 6$:

\begin{equation}
D_E=\frac{1}{16}\sqrt{\frac{\pi k_B T}{m}}\sqrt[3]{\frac{6}{\pi
\rho \varphi ^2}} {\frac{(1-\varphi)^3}{(1-\varphi/2)}} \label{DE}
\end{equation}


Summing up, by virtue of the results recalled sofar, the
$S(Q,\omega)$ in dense fluids, such as in the present case of a
monoatomic liquid close to the melting point, can be described to
a sufficient extent of accuracy in terms of three Lorentzian lines
(effective hydrodynamic modes) up to relatively large wavevectors
($Q\approx 20$ nm$^{-1}$), while above this value the acoustic
mode is overdamped and the quasi-elastic extended heat mode should
dominate ($Ql_E>>0.05$). The full width at half maximum (FWHM) of
the quasi-elastic mode is quantitatively predicted by
Eq.s~(\ref{zh_hs}), (\ref{dq}) and (\ref{DE}).

\begin{figure} [h]
\centering
\includegraphics[width=.5\textwidth]{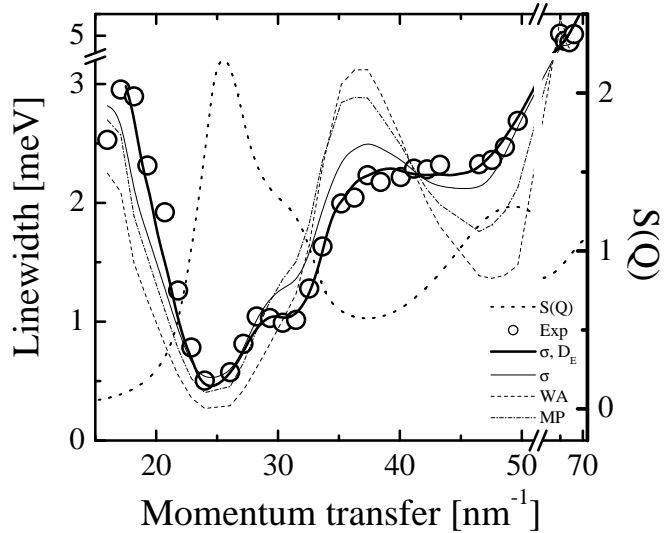}
\vspace{-6.cm}\caption{FWHM of the dynamic structure factor (left
axis): experimental determination (open circles, error bars
smaller than the symbol size) and different predictions according
to the revised Enskog's theory (Eq.(\ref{zh_hs})). Thin continuous
line: $\sigma$ is the only fitting parameter ($D_E$ is determined
exploiting Eq.(\ref{DE})). Thick continuous line: both $\sigma$
and $D_E$ are fitting parameters. Thin dashed line (WA): $\sigma$
is determined as the value corresponding to the hard sphere
structure factor which better describes the measured $S(Q)$ ($D_E$
is the only fitting parameter). Thin dot-dashed line (MP):
$\sigma$ is determined from the main peak of the structure factor,
through Eq.(\ref{qm}) ($D_E$ is the only fitting parameter). Also
reported is the static structure factor (dotted line, right axis),
which drives the oscillations in the FWHM.} \label{width}
\end{figure}

The half width at half maximum of the quasi-elastic line (heat
mode) are reported in Fig. \ref{width} as a function of the $Q$
values, together with the $S(Q)$ \cite{bf}. Here, one can clearly
observe oscillations in the line-width driven by the $S(Q)$. The
second minimum at $Q\approx 30$ nm$^{-1}$, in particular, is
clearly related to the shoulder of the $S(Q)$ observed the same
$Q$ position. The line-width predicted by Eq.(\ref{zh_hs}) is also
reported, with different choices of the two parameters $\sigma$
and $D_E$ (see table \ref{table}). By leaving both parameters free
one gets a best fit value of $\sigma =0.279$ nm, corresponding to
a packing fraction of $\varphi=0.601$ (and a reduced density
$V_0/V=0.811$) which is just beyond the maximum theoretical value
for hard spheres ($\varphi=0.545$). This seems to be a reasonable
result for a dense liquid, like the one under investigation here,
recalling that $\sigma$ has to be regarded to as an effective
parameter, thus not necessarily \textit{strictly} related to the
density through Eq. (\ref{reddens}). Using only $\sigma$ as free
parameter (i.e. exploiting Eq.(\ref{DE})) one still gets
reasonable agreement (thin continuous line in Fig.\ref{width}),
even in the region of the secondary shoulder.

Aiming at a description of microscopic properties of a simple
fluid within the hard sphere paradigm, alternative choices of the
effective particle diameter can be considered. One can, indeed,
choose $\sigma$ from the position of the first peak $Q_M$ of the
structure factor \cite{coh_hs} (MP):

\begin{equation}
\sigma=\frac{2 \pi}{Q_M} \label{qm}
\end{equation}

or, through Eq.(\ref{reddens}), by adjusting the value of the
reduced density to maximize the whole $S(Q)$ agreement
\cite{tsay_ga} (WA). We tried both, either using Eq.(\ref{DE}) or
not, and the agreement with the experimental line-width is poor.
The results are quantitatively summarized in table \ref{table}: by
fixing the hard- sphere diameter with one of the two mentioned
criteria the $\chi ^2$ is definitely worse. Moreover, looking at
Fig.(\ref{width}), one clearly observes that with this choices the
secondary minimum observed in the FWHM at $Q\approx 30$ nm$^{-1}$
is smeared out. This feature, therefore, is not the mere
consequence of the oscillations in $S(Q)$ (De Gennes narrowing).
It testifies to the presence of hard sphere dynamics and allows
the sharp determination of an effective hard sphere diameter via
the Bessel terms of Eq.(\ref{zh_hs}).

But what is the physical meaning of such "dynamical" effective
diameter, and why it is larger than those obtained from the static
structure factor? As we already mentioned, a description in terms
of a uniform distribution of hard spheres does not apply to liquid
gallium, as well as to several other liquid metals (Zn, Cd, Bi,
Si). The covalency residue of Ga, in particular, has been
rationalized in terms of dimer-like structures \cite{gong_ga}, and
clustering effects have been hypotized aiming to reproduce the
$S(Q)$ \cite{tsay_ga}. The effective diameter that we find,
therefore, could be an indication of the supra-atomic nature of
the "effective particles". Consistently, plugging our effective
$\sigma =0.279$ nm in Eq. (\ref{DE}), and solving it for a
correspondent effective mass, we get $m_{eff}=83$ a.m.u., which is
larger than the atomic mass of gallium $m=69.7$ a.m.u. The
effective mass and diameter, therefore, could be regarded as the
mean sizes of the "clusters" undergoing collective dynamics. This
hypothesis, however, needs to be tested extending the present
study to other systems and at different temperature, as the
structural anomalies are known to be strongly temperature
dependent.

In summary, we have shown how the microscopic dynamics of a well
known non hard-sphere liquid, namely liquid Gallium at the melting
point, is surprisingly well described by Enskog's theory,
analytically derived for a hard-sphere fluid. More specifically,
the connection between structure and dynamics, which is the main
outcome of this theory, is robust enough to persist even in a
wavelength region where anomalies typical of non simple liquids
develop. From the experimental data, therefore, an
\textit{effective} hard-sphere diameter can be obtained, which
turns out to be larger than the one related to structural
properties according to well established methods. This result
suggests, therefore, an extended validity of the revised Enskog's
theory beyond the class of systems (hard spheres) for which it was
derived, and provides an experimental route to the determination
of an \textit{effective dynamical} diameter.

The authors are grateful to P. Ascarelli, E. G. D. Cohen, F.
Sciortino and J.-B. Suck for fruitful discussions. The synchrotron
radiation experiment was performed at the SPring-8 with the
approval of the Japan Synchrotron Radiation Research Institute
(JASRI) (Proposal No. 2004A0634-ND3d-np).


\begin{thebibliography}{22}
\expandafter\ifx\csname
natexlab\endcsname\relax\def\natexlab#1{#1}\fi
\expandafter\ifx\csname bibnamefont\endcsname\relax
  \def\bibnamefont#1{#1}\fi
\expandafter\ifx\csname bibfnamefont\endcsname\relax
  \def\bibfnamefont#1{#1}\fi
\expandafter\ifx\csname citenamefont\endcsname\relax
  \def\citenamefont#1{#1}\fi
\expandafter\ifx\csname url\endcsname\relax
  \def\url#1{\texttt{#1}}\fi
\expandafter\ifx\csname
urlprefix\endcsname\relax\def\urlprefix{URL }\fi
\providecommand{\bibinfo}[2]{#2}
\providecommand{\eprint}[2][]{\url{#2}}

\bibitem[{\citenamefont{Chapman and Cowling}(1952)}]{ens_hs}
\bibinfo{author}{\bibfnamefont{S.}~\bibnamefont{Chapman}} \bibnamefont{and}
  \bibinfo{author}{\bibfnamefont{T.}~\bibnamefont{Cowling}},
  \emph{\bibinfo{title}{The mathematical theory of non uniform gases}}
  (\bibinfo{year}{1952}).

\bibitem[{\citenamefont{Lebowitz et~al.}(1969)\citenamefont{Lebowitz, Percus,
  and Sykes}}]{lebo_hs}
\bibinfo{author}{\bibfnamefont{J.~L.} \bibnamefont{Lebowitz}},
  \bibinfo{author}{\bibfnamefont{J.~K.} \bibnamefont{Percus}},
  \bibnamefont{and} \bibinfo{author}{\bibfnamefont{J.}~\bibnamefont{Sykes}},
  \bibinfo{journal}{Phys. Rev.} \textbf{\bibinfo{volume}{188}},
  \bibinfo{pages}{487} (\bibinfo{year}{1969}).

\bibitem[{\citenamefont{Furtado et~al.}(1975)\citenamefont{Furtado, Mazenko,
  and Yip}}]{furt_hs}
\bibinfo{author}{\bibfnamefont{P.~M.} \bibnamefont{Furtado}},
  \bibinfo{author}{\bibfnamefont{G.~F.} \bibnamefont{Mazenko}},
  \bibnamefont{and} \bibinfo{author}{\bibfnamefont{S.}~\bibnamefont{Yip}},
  \bibinfo{journal}{Phys. Rev. A} \textbf{\bibinfo{volume}{12}},
  \bibinfo{pages}{1653} (\bibinfo{year}{1975}).

\bibitem[{\citenamefont{Alley et~al.}(1983)\citenamefont{Alley, Alder, and
  Yip}}]{all_hs1}
\bibinfo{author}{\bibfnamefont{W.~E.} \bibnamefont{Alley}},
  \bibinfo{author}{\bibfnamefont{B.~J.} \bibnamefont{Alder}}, \bibnamefont{and}
  \bibinfo{author}{\bibfnamefont{S.}~\bibnamefont{Yip}},
  \bibinfo{journal}{Phys. Rev. A} \textbf{\bibinfo{volume}{27}},
  \bibinfo{pages}{3174} (\bibinfo{year}{1983}).

\bibitem[{\citenamefont{Alley and Alder}(1983)}]{all_hs}
\bibinfo{author}{\bibfnamefont{W.~E.} \bibnamefont{Alley}} \bibnamefont{and}
  \bibinfo{author}{\bibfnamefont{B.~J.} \bibnamefont{Alder}},
  \bibinfo{journal}{Phys. Rev. A} \textbf{\bibinfo{volume}{27}},
  \bibinfo{pages}{3158} (\bibinfo{year}{1983}).

\bibitem[{\citenamefont{{de Schepper} and Cohen}(1980)}]{desh_hyd0}
\bibinfo{author}{\bibfnamefont{I.~M.} \bibnamefont{{de Schepper}}}
  \bibnamefont{and} \bibinfo{author}{\bibfnamefont{E.~G.~D.}
  \bibnamefont{Cohen}}, \bibinfo{journal}{Phys. Rev. A}
  \textbf{\bibinfo{volume}{22}}, \bibinfo{pages}{287} (\bibinfo{year}{1980}).

\bibitem[{\citenamefont{{de Schepper} et~al.}(1984)\citenamefont{{de Schepper},
  Cohen, and Zuilhof}}]{desh_hs}
\bibinfo{author}{\bibfnamefont{I.~M.} \bibnamefont{{de Schepper}}},
  \bibinfo{author}{\bibfnamefont{E.~G.~D.} \bibnamefont{Cohen}},
  \bibnamefont{and} \bibinfo{author}{\bibfnamefont{M.~J.}
  \bibnamefont{Zuilhof}}, \bibinfo{journal}{Phys. Lett.}
  \textbf{\bibinfo{volume}{101A}}, \bibinfo{pages}{399} (\bibinfo{year}{1984}).

\bibitem[{\citenamefont{Cohen et~al.}(1984)\citenamefont{Cohen, {de Schepper},
  and Zuilhof}}]{coh_hs1}
\bibinfo{author}{\bibfnamefont{E.~G.~D.} \bibnamefont{Cohen}},
  \bibinfo{author}{\bibfnamefont{I.~M.} \bibnamefont{{de Schepper}}},
  \bibnamefont{and} \bibinfo{author}{\bibfnamefont{M.~J.}
  \bibnamefont{Zuilhof}}, \bibinfo{journal}{Physica}
  \textbf{\bibinfo{volume}{127B}}, \bibinfo{pages}{282} (\bibinfo{year}{1984}).

\bibitem[{\citenamefont{{Kamgar-Parsi}
  et~al.}(1987)\citenamefont{{Kamgar-Parsi}, Cohen, and {de
  Schepper}}}]{kag_hs}
\bibinfo{author}{\bibfnamefont{B.}~\bibnamefont{{Kamgar-Parsi}}},
  \bibinfo{author}{\bibfnamefont{E.~G.~D.} \bibnamefont{Cohen}},
  \bibnamefont{and} \bibinfo{author}{\bibfnamefont{I.~M.} \bibnamefont{{de
  Schepper}}}, \bibinfo{journal}{Phys. Rev. A} \textbf{\bibinfo{volume}{35}},
  \bibinfo{pages}{4781} (\bibinfo{year}{1987}).

\bibitem[{\citenamefont{Cohen et~al.}(1987)\citenamefont{Cohen, Westerhuijs,
  and {de Schepper}}}]{coh_hs}
\bibinfo{author}{\bibfnamefont{E.~G.~D.} \bibnamefont{Cohen}},
  \bibinfo{author}{\bibfnamefont{P.}~\bibnamefont{Westerhuijs}},
  \bibnamefont{and} \bibinfo{author}{\bibfnamefont{I.~M.} \bibnamefont{{de
  Schepper}}}, \bibinfo{journal}{Phys. Rev. Lett.}
  \textbf{\bibinfo{volume}{59}}, \bibinfo{pages}{2872} (\bibinfo{year}{1987}).

\bibitem[{\citenamefont{Dahlborg and Olsson}(1982)}]{dal_bi}
\bibinfo{author}{\bibfnamefont{U.}~\bibnamefont{Dahlborg}} \bibnamefont{and}
  \bibinfo{author}{\bibfnamefont{L.~G.} \bibnamefont{Olsson}},
  \bibinfo{journal}{Phys. Rev. A} \textbf{\bibinfo{volume}{25}},
  \bibinfo{pages}{2712} (\bibinfo{year}{1982}).

\bibitem[{\citenamefont{Gong et~al.}(1991)\citenamefont{Gong, Chiarotti,
  Parrinello, and Tosatti}}]{gong_ga}
\bibinfo{author}{\bibfnamefont{X.~G.} \bibnamefont{Gong}},
  \bibinfo{author}{\bibfnamefont{G.~L.} \bibnamefont{Chiarotti}},
  \bibinfo{author}{\bibfnamefont{M.}~\bibnamefont{Parrinello}},
  \bibnamefont{and} \bibinfo{author}{\bibfnamefont{E.}~\bibnamefont{Tosatti}},
  \bibinfo{journal}{Phys. Rev. B} \textbf{\bibinfo{volume}{43}},
  \bibinfo{pages}{R14277} (\bibinfo{year}{1991}).

\bibitem[{\citenamefont{Ascarelli}(1966)}]{asc_ga}
\bibinfo{author}{\bibfnamefont{P.}~\bibnamefont{Ascarelli}},
  \bibinfo{journal}{Phys. Rev. B} \textbf{\bibinfo{volume}{143}},
  \bibinfo{pages}{36} (\bibinfo{year}{1966}).

\bibitem[{\citenamefont{Bellissent-Funel
  et~al.}(1989)\citenamefont{Bellissent-Funel, Chieux, Levesque, and
  Weis}}]{bf}
\bibinfo{author}{\bibfnamefont{M.~C.} \bibnamefont{Bellissent-Funel}},
  \bibinfo{author}{\bibfnamefont{P.}~\bibnamefont{Chieux}},
  \bibinfo{author}{\bibfnamefont{D.}~\bibnamefont{Levesque}}, \bibnamefont{and}
  \bibinfo{author}{\bibfnamefont{J.~J.} \bibnamefont{Weis}},
  \bibinfo{journal}{Phys. Rev. A} \textbf{\bibinfo{volume}{39}},
  \bibinfo{pages}{6310} (\bibinfo{year}{1989}).

\bibitem[{\citenamefont{Bermejo et~al.}(1994)\citenamefont{Bermejo,
  Garcia-Hernandez, Martinez, and Hennion}}]{ber_ga1}
\bibinfo{author}{\bibfnamefont{F.~J.} \bibnamefont{Bermejo}},
  \bibinfo{author}{\bibfnamefont{M.}~\bibnamefont{Garcia-Hernandez}},
  \bibinfo{author}{\bibfnamefont{J.~L.} \bibnamefont{Martinez}},
  \bibnamefont{and} \bibinfo{author}{\bibfnamefont{B.}~\bibnamefont{Hennion}},
  \bibinfo{journal}{Phys. Rev. E} \textbf{\bibinfo{volume}{49}},
  \bibinfo{pages}{3133} (\bibinfo{year}{1994}).

\bibitem[{\citenamefont{Bermejo et~al.}(1997)\citenamefont{Bermejo,
  Fernandez-Perea, Alvarez, Roessli, Fischer, and Bossy}}]{ber_ga2}
\bibinfo{author}{\bibfnamefont{F.~J.} \bibnamefont{Bermejo}},
  \bibinfo{author}{\bibfnamefont{R.}~\bibnamefont{Fernandez-Perea}},
  \bibinfo{author}{\bibfnamefont{M.}~\bibnamefont{Alvarez}},
  \bibinfo{author}{\bibfnamefont{B.}~\bibnamefont{Roessli}},
  \bibinfo{author}{\bibfnamefont{H.~E.} \bibnamefont{Fischer}},
  \bibnamefont{and} \bibinfo{author}{\bibfnamefont{J.}~\bibnamefont{Bossy}},
  \bibinfo{journal}{Phys. Rev. E} \textbf{\bibinfo{volume}{56}},
  \bibinfo{pages}{3358} (\bibinfo{year}{1997}).

\bibitem[{\citenamefont{Bove et~al.}(2005)\citenamefont{Bove, Formisano,
  Sacchetti, Petrillo, Ivanov, Dorner, and Barocchi}}]{bov_ga}
\bibinfo{author}{\bibfnamefont{L.~E.} \bibnamefont{Bove}},
  \bibinfo{author}{\bibfnamefont{F.}~\bibnamefont{Formisano}},
  \bibinfo{author}{\bibfnamefont{F.}~\bibnamefont{Sacchetti}},
  \bibinfo{author}{\bibfnamefont{C.}~\bibnamefont{Petrillo}},
  \bibinfo{author}{\bibfnamefont{A.}~\bibnamefont{Ivanov}},
  \bibinfo{author}{\bibfnamefont{B.}~\bibnamefont{Dorner}}, \bibnamefont{and}
  \bibinfo{author}{\bibfnamefont{F.}~\bibnamefont{Barocchi}},
  \bibinfo{journal}{Phys. Rev. B} \textbf{\bibinfo{volume}{71}},
  \bibinfo{pages}{014207} (\bibinfo{year}{2005}).

\bibitem[{\citenamefont{Scopigno et~al.}(2002)\citenamefont{Scopigno,
  Filipponi, Krisch, Monaco, Ruocco, and Sette}}]{scop_prlga}
\bibinfo{author}{\bibfnamefont{T.}~\bibnamefont{Scopigno}},
  \bibinfo{author}{\bibfnamefont{A.}~\bibnamefont{Filipponi}},
  \bibinfo{author}{\bibfnamefont{M.}~\bibnamefont{Krisch}},
  \bibinfo{author}{\bibfnamefont{G.}~\bibnamefont{Monaco}},
  \bibinfo{author}{\bibfnamefont{G.}~\bibnamefont{Ruocco}}, \bibnamefont{and}
  \bibinfo{author}{\bibfnamefont{F.}~\bibnamefont{Sette}},
  \bibinfo{journal}{Phys. Rev. Lett.} \textbf{\bibinfo{volume}{89}},
  \bibinfo{pages}{255506} (\bibinfo{year}{2002}).

\bibitem[{\citenamefont{Baron et~al.}(2000)\citenamefont{Baron, Tanaka, Goto,
  Takeshita, Matsushita, and Ishikawa}}]{bar_spring8}
\bibinfo{author}{\bibfnamefont{A.~Q.~R.} \bibnamefont{Baron}},
  \bibinfo{author}{\bibfnamefont{Y.}~\bibnamefont{Tanaka}},
  \bibinfo{author}{\bibfnamefont{S.}~\bibnamefont{Goto}},
  \bibinfo{author}{\bibfnamefont{K.}~\bibnamefont{Takeshita}},
  \bibinfo{author}{\bibfnamefont{T.}~\bibnamefont{Matsushita}},
  \bibnamefont{and} \bibinfo{author}{\bibfnamefont{T.}~\bibnamefont{Ishikawa}},
  \bibinfo{journal}{J. Phys. Chem. Solids} \textbf{\bibinfo{volume}{61}},
  \bibinfo{pages}{461} (\bibinfo{year}{2000}).

\bibitem[{\citenamefont{Scopigno et~al.}(2000)\citenamefont{Scopigno, Balucani,
  Ruocco, and Sette}}]{scop_jpc}
\bibinfo{author}{\bibfnamefont{T.}~\bibnamefont{Scopigno}},
  \bibinfo{author}{\bibfnamefont{U.}~\bibnamefont{Balucani}},
  \bibinfo{author}{\bibfnamefont{G.}~\bibnamefont{Ruocco}}, \bibnamefont{and}
  \bibinfo{author}{\bibfnamefont{F.}~\bibnamefont{Sette}}, \bibinfo{journal}{J.
  Phys. C} \textbf{\bibinfo{volume}{12}}, \bibinfo{pages}{8009}
  (\bibinfo{year}{2000}).

\bibitem[{\citenamefont{Faber}(1972)}]{FABER}
\bibinfo{author}{\bibfnamefont{T.}~\bibnamefont{Faber}},
  \emph{\bibinfo{title}{Introduction to the Theory of Liquid Metals}}
  (\bibinfo{publisher}{Cambridge University Press, Cambridge},
  \bibinfo{year}{1972}).

\bibitem[{\citenamefont{Tsay and Wang}(1994)}]{tsay_ga}
\bibinfo{author}{\bibfnamefont{S.~F.} \bibnamefont{Tsay}} \bibnamefont{and}
  \bibinfo{author}{\bibfnamefont{S.}~\bibnamefont{Wang}},
  \bibinfo{journal}{Phys. Rev. B} \textbf{\bibinfo{volume}{50}},
  \bibinfo{pages}{108} (\bibinfo{year}{1994}).

\end{thebibliography}

\end{document}